\documentclass[12pt,twocolumn]{iopart}
 \expandafter\let\csname equation*\endcsname\relax
  \expandafter\let\csname endequation*\endcsname\relax

\usepackage{amsmath,amssymb}
\usepackage{dblfloatfix}
\usepackage[version=3]{mhchem}
\usepackage{graphicx}
\usepackage{booktabs}
\usepackage{subfig}

\begin{document}

\title{Single-Crystal Growth and Thermoelectric Properties of \ce{Ge(Bi $,$ Sb)_{4}Te_7}}

\author{Fabian von Rohr$^{1,2}$, Andreas Schilling$^{1}$, Robert J. Cava$^{2}$}

\address{$^{1}$ Physik-Institut, Universit\"at Z\"urich, Winterthurerstrasse 190, CH-8057 Z\"urich, Switzerland \\
$^{2}$ Department of Chemistry, Princeton University, Princeton, New Jersey 08544, USA}
\ead{vonrohr@physik.uzh.ch}
\begin{abstract}

The thermoelectric properties between 10 and 300 K and the growth of single crystals of \textit{n}-type and \textit{p}-type \ce{GeBi4Te7}, \ce{GeSb4Te7}, and the \ce{Ge(Bi_{1-\textit{x}}Sb_{\textit{x}})_{4}Te7} solid solution are reported. Single crystals were grown by the modified Bridgman method, and \textit{p}-type behavior was achieved by the substitution of Bi by Sb in \ce{GeBi4Te7}. The thermopower in the \ce{Ge(Bi_{1-\textit{x}}Sb_{\textit{x}})_4Te7} solid solution ranges from $\mathrm{-117 \ \mu V K^{-1}}$ to $\mathrm{+160 \ \mu V K^{-1}}$. The crossover from \textit{n}-type to \textit{p}-type is continuous with increasing Sb content and is observed at $x \ \mathrm{ \approx \ 0.15}$. The highest thermoelectric efficiencies  among the tested \textit{n}-type and \textit{p}-type samples are $Z_{n}T \mathrm{\ = \ 0.11}$ and $Z_{p}T \mathrm{ \ = \ 0.20}$, respectively. For an optimal \textit{n}-\textit{p} couple in this alloy system the composite figure of merit is $Z_{np}T \mathrm{ \ = \ 0.17}$  at room temperature.

\end{abstract}

\maketitle

\section{Introduction}
Thermoelectric materials exhibit significant potential for technological applications, e.g. efficient cooling below 300 K and waste-heat recovery at elevated temperatures. The unusual transport properties of these materials have been subject of extensive studies. (see, e.g. \cite{thermonature}) \ce{Bi2Te3}-based solid solutions have been investigated in the past as excellent thermoelectric materials for room temperature applications. \cite{ThuH} However, the closely related layered ternary compounds with the general pseudobinary compositions $n\mathrm{ATe}-m\mathrm{B_2Te_3}$ (with A = Ge, Pb, Sn and B = Bi, Sb; and \textit{n},\textit{m} = integers) have received little attention as potential thermoelectric materials, despite reported large thermopowers \cite{nonstoichiometry} \cite{Shelimova} and their expected low lattice thermal conductivities due to their large, complex unit cells. \cite{Kuznetsova1} \ce{AB_4Te_7} (\textit{n}=1, \textit{m}=2) compounds crystallize in the hexagonal space group $P\overline{3}m1$. \cite{GBT147_disc} The crystal structures consist of stacked - [Te-B-Te-B-Te]-[ Te-B-Te-A-Te-B-Te]-units. A 12-layer stack, including two Te-Te van der Waals gaps as preferred cleavage planes, forms the unit cell. The bonding inside the multi-layered packets has an ionic-covalent character. \\

In addition to their potential as thermoelectrics, these materials are also candidates for "topological insulators" - materials with a bulk energy gap but topologically protected metallic surface states. \cite{Kane} Experiments have characterized charge transport of these surface electrons  in \ce{Bi_2Te_3} and related compounds. \cite{Cava_transport} The discovery of metallic surface states energetically well separated from the bulk states in \ce{GeBi4Te7} \cite{GBT_dirac_fvr}, reinforces the need for the growth of high-quality samples of this compound. Here, we report the single crystal growth of this phase and the bulk transport properties from 10 K to 300 K of the  \ce{Ge(Bi$,$Sb)_4Te7} solid solution with the aim of characterizing its potential as a thermoelectric material. The current work builds on the preliminary characterization of this system reported previously. \cite{Shelimova} The results indicate future pathways to optimization of the compound as both a thermoelectric and a topological insulator. 

\section{Experiment}
Stoichiometric mixtures of Ge (99.99\%), Te (99.99\%), Sb (99.999\%) and Bi (99.999\%) were heated to 950 $\mathrm{^{\circ}C}$ for 10 hours in 12 mm diameter quartz tubes in vacuum. The best crystals were obtained by cooling to 450  $\mathrm{^{\circ}C}$ at 50  $\mathrm{^{\circ}C/h}$ and then annealing at that temperature for one week. In a final step, the crystals were quenched into water. The single crystals were indentified to be single phase by X-ray powder diffraction using a Stoe STADIP diffractometer (Cu-$\mathrm{K}_{\alpha\mathrm{1}}$ radiation, $\lambda$ = 1.54051 \AA, Ge-monochromator). Single crystals were easily cleaved along the \textit{c}-plane at the Te-Te van der Waals gaps. Transport measurements were performed on well shaped polycrystalline samples (crushed and pelletized single crystals), which were additionally annealed at 450 $\mathrm{^{\circ}C}$ for 24 h and on single crystals with the current in the $c$-plane. Temperature dependent measurements of the thermopower as well as the electrical and thermal transport properties were carried out in a Quantum Design Physical Property Measurement System (PPMS) equipped with the Thermal Transport Option (TTO). Thermopower measurements were also performed using a homemade probe with an MMR Technologies SB100 Seebeck measurement system. 

\section{Results and Discussion}
\begin{figure}[p]
\centering
\includegraphics [width=\textwidth] {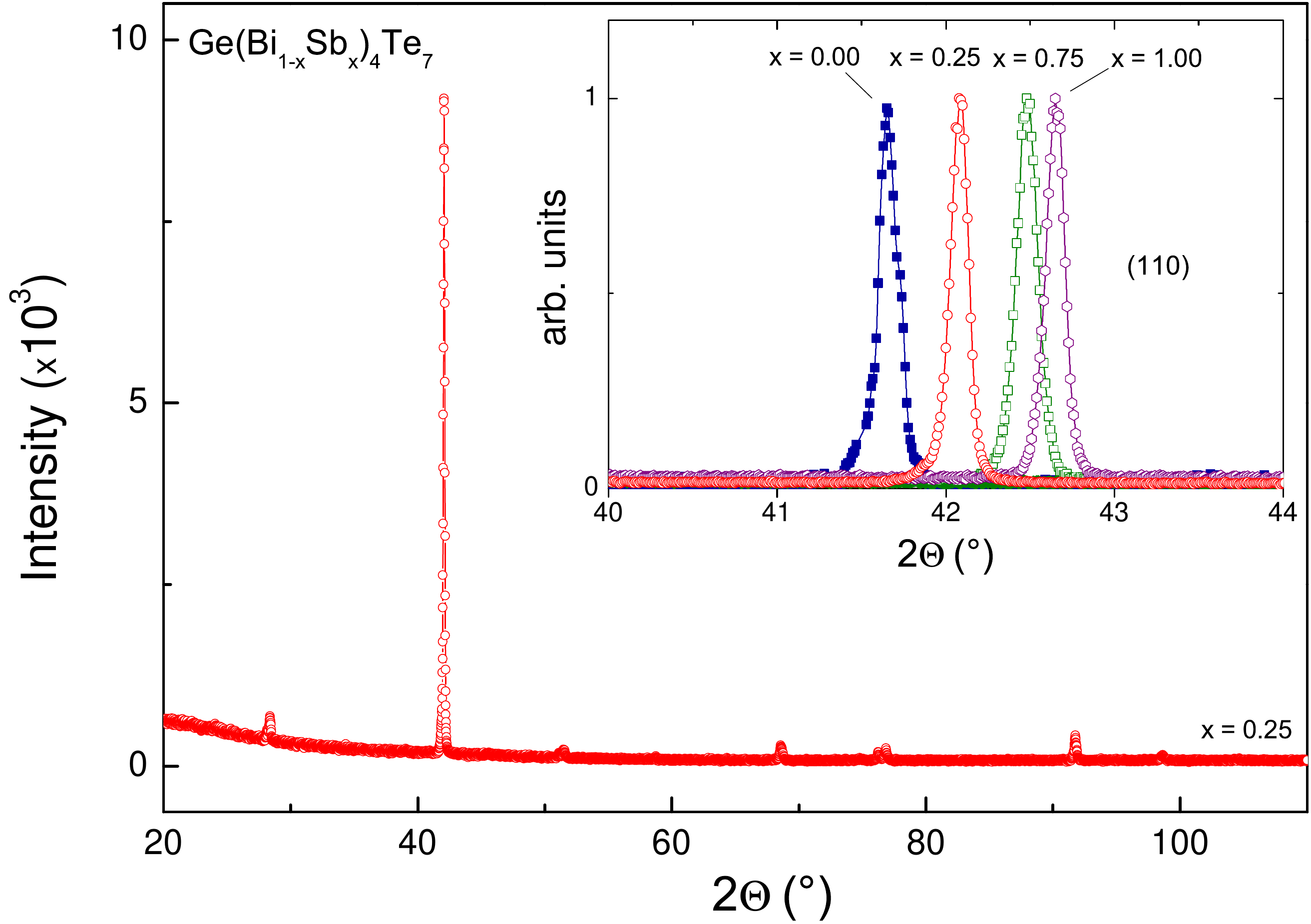}
\caption{The X-ray diffraction pattern of \textit{x} = 0.25 showing the reflections from the basal plane of the cleaved single crystal. In the inset we show the change of the (110) reflection for \textit{x} = 0, 0.25, 0.75 and 1.}
\label{fig:xray}
\end{figure}

Powder X-ray diffraction patterns of all samples of the \ce{Ge(Bi_{1-\textit{x}}Sb_{\textit{x}})_4Te_7} solid solution were indexed using the hexagonal space group $P\overline{3}m1$. The XRD pattern for \textit{x} = 0.25 and the change within the series of the (110) reflection (inset of figure \ref{fig:xray}) as a function of doping for \textit{x} = 0, 0.25, 0.75 and 1 are shown in figure \ref{fig:xray}. 
The crystallites in the X-ray diffraction patterns display preferred orientation, as is expected for strongly layered compounds. Replacing Bi by Sb leads to a decrease of the unit cell volume due to the smaller ionic radius of Sb. Within the series, the lattice parameters vary from approximately \textit{a} = 4.29 \AA, \textit{c} = 23.89 \AA \ for the pure Bi end member to \textit{a} = 4.24 \AA, \textit{c} = 23.80 \AA \ for the pure Sb end member.  \\

\begin{figure}[p]
\centering
\includegraphics [width=\textwidth] {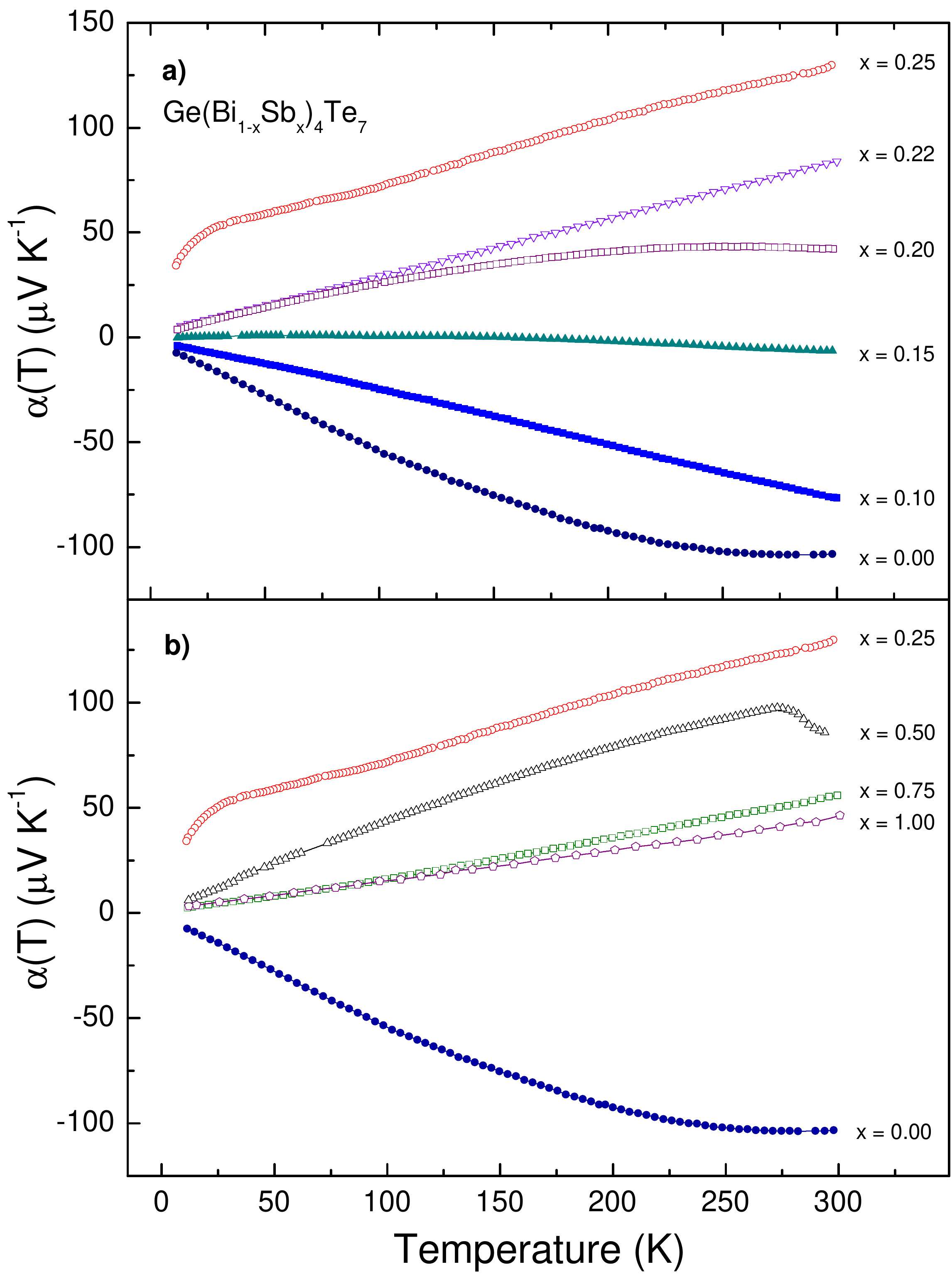}
\caption{Upper panel - Temperature dependence of the thermopower for polycrystalline samples of \textit{x} = 0.0, 0.1, 0.15, 0.22 and 0.25 in a temperature range from 10 K to 300 K. The crossover from a negative to a positive thermopower as a function of Sb-doping is observed at $x \mathrm{ \ \approx \ 0.15}$. In the lower panel we show the temperature dependent thermopowers for \textit{x} = 0, 0.25, 0.5, 0.75 and 1.}

\label{fig:seebeck}
\end{figure}

The temperature dependent measurements of the thermopower $\mathrm{\alpha(T)}$ in the range from 10 K to 300 K for \textit{x} = 0, 0.1, 0.15, 0.22 and 0.25 are presented in figure \ref{fig:seebeck}a; these compositions are selected to illustrate the \textit{n}- to \textit{p}- crossover. Within the solid solution of polycrystalline \ce{Ge(Bi $,$ Sb)_{4}Te_7} thermopowers ranging from -104 $\mathrm{\mu V K^{-1}}$ to +130 $\mathrm{\mu V K^{-1}}$ are found. Nominally stoichiometric \ce{GeBi_4Te_7} has the highest negative thermopower within the series, and is therefore strongly \textit{n}-type. Upon substitution of Bi by Sb, the thermopower decreases gradually for \textit{x} = 0.1 and \textit{x} = 0.15, and eventually crosses to a positive thermopower around $x \mathrm{ \ \approx \ 0.15}$. The highest positive thermopower is observed for \textit{x} = 0.25. In figure \ref{fig:seebeck}b we present the overall behavior of the full series, using representative compositions between \textit{x} = 0 and \textit{x} = 1. \\

\begin{figure}[p]
\centering
\includegraphics [width=\textwidth] {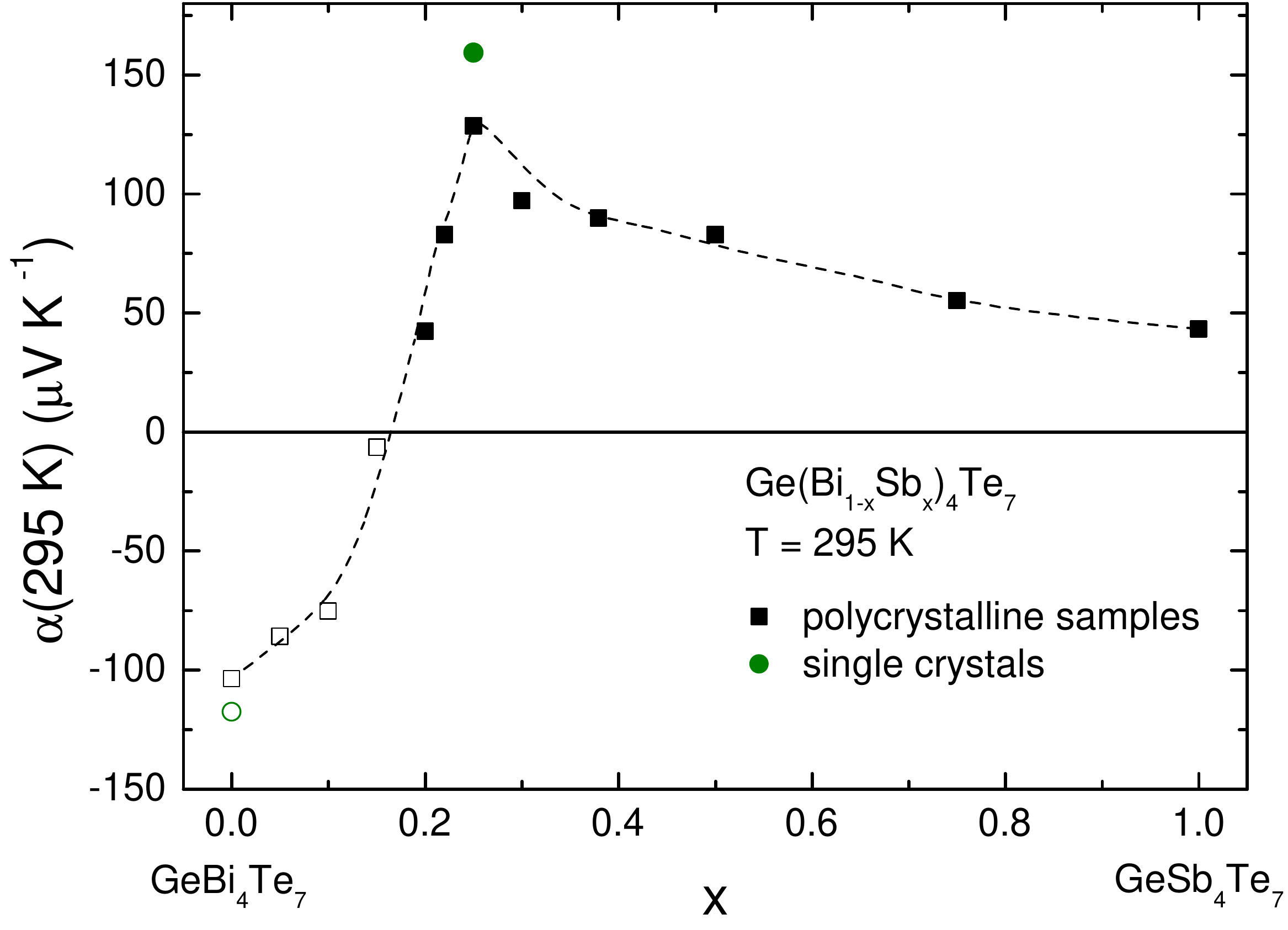}
\caption{Composition dependence of the room temperature thermopowers of \ce{Ge(Bi_{1-\textit{x}}Sb_{\textit{x}})_4Te_7} for \textit{x} ranging from 0 to 1, for 12 differently doped polycrystalline samples, and two single crystals (x = 0 and 0.25). The dashed line is a guide to the eye for the room temperature thermopowers of the polycrystalline samples.}
\label{fig:seebeck295}
\end{figure}

In figure \ref{fig:seebeck295} the room temperature thermopowers $\mathrm{\alpha_{295 K}}$ of \ce{Ge(Bi_{1-\textit{x}}Sb_{\textit{x}})_4Te_7} for the whole range from \textit{x} = 0 to \textit{x} = 1, measured for 12 differently doped polycrystalline samples and two single crystals are shown. For Sb contents higher than \textit{x} = 0.25, the thermopower decreases gradually until it reaches for the end member of the series, \ce{GeSb4Te7}, a value of +43 $\mu V K^{-1}$ at room temperature. The observed behavior of the thermopower is typical for an \textit{n}-type to \textit{p}-type crossover. The highest positive thermopower is observed in the vicinity of the transition from \textit{n}-type to \textit{p}-type, similar to what is observed in tetradymite-based thermoelectrics. (e.g. \cite{Mike}) \\

\begin{figure*} [p]
\centering
{\includegraphics[width=\textwidth]{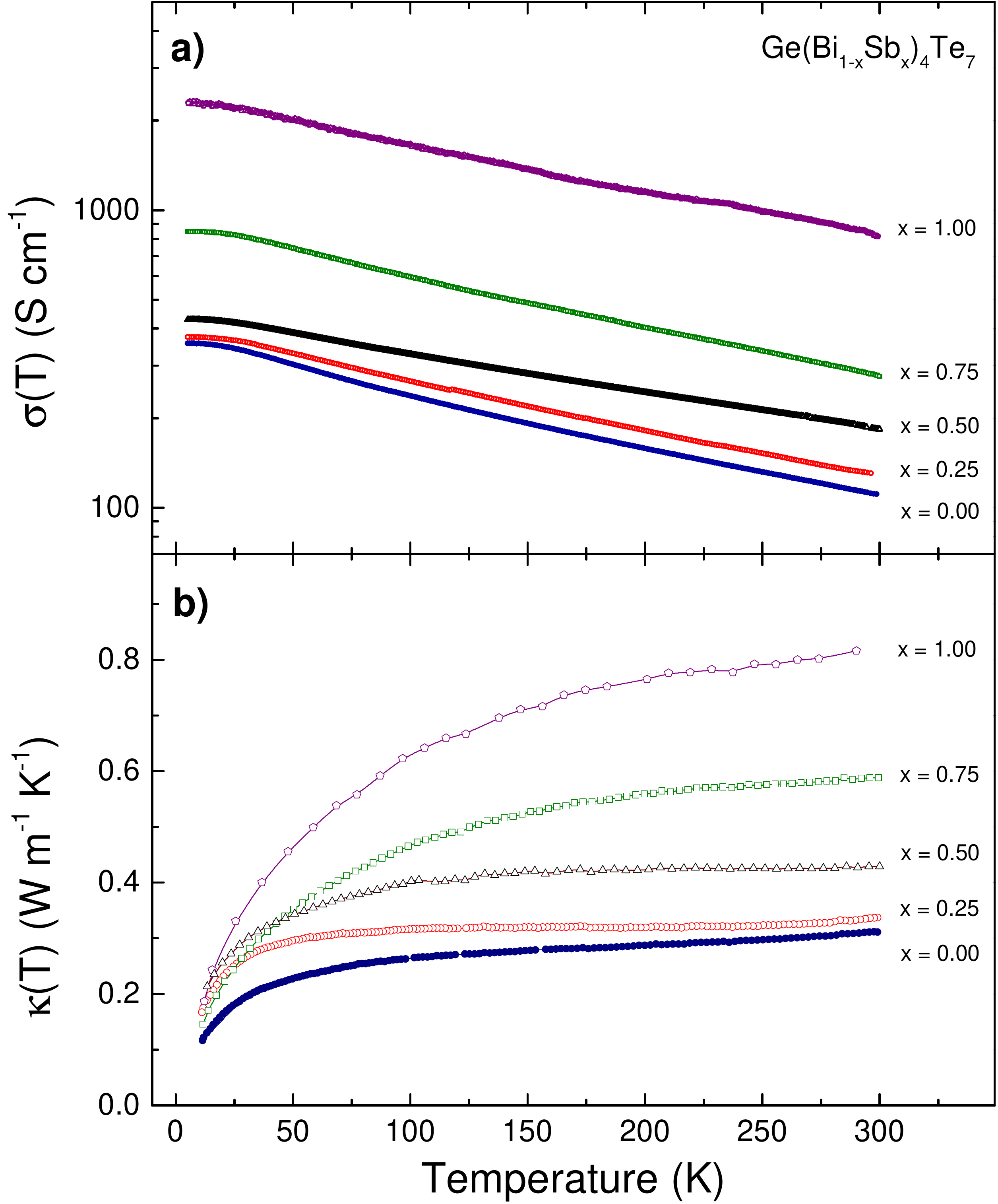}}
\caption{(a) Temperature dependent electrical conductivity $\sigma(T)$ for \textit{x} = 0, 0.25, 0.5, 0.75 and 1. (b) The temperature dependent thermal conductivity $\kappa(T)$ for \textit{x} = 0, 0.25, 0.5, 0.75, and 1.}
\label{fig:restherm}
\end{figure*}

The temperature dependent electrical conductivity measurements $\sigma(T)$ are shown in figure \ref{fig:restherm}a. All samples show the semimetallic behavior that is typical for heavily doped semiconductors, with electrical conductivities $\sigma(T)$ in the range of 150 to 2200 $\mathrm{S \ cm^{-1}}$ ($\rho \approx$ 1 to 10 $\mathrm{m\Omega \ cm}$). The crossover from \textit{n}-type to \textit{p}-type is not accompanied by a semiconducting or insulating resistivity regime. As a general trend we observe that the electrical conductivity increases for increasing Sb contents. In figure \ref{fig:restherm}b we show the temperature dependent thermal conductivities $\kappa(T)$. They range from 0.2 $\mathrm{W K^{-1} m^{-1}}$ for \textit{x} = 0 to 0.8 $\mathrm{W K^{-1} m^{-1}}$ for \textit{x} = 1. The thermal conductivities $\kappa(T)$ increase as a function of Sb content. They are considerably lower than for optimized $\mathrm{(Bi,Sb)_2(Te,Se)_3}$ \cite{ThuH}, presumably due to both the increased complexity of the unit cell and the phonon scattering caused by the likely (Bi, Ge,  Sb) atomic site disorder.\\

\begin{equation}
\frac{\kappa_e}{\sigma} = \frac{\pi^2}{3} \left(\frac{k_b}{e}\right)^2 T = L T
\label{eq1} 
\end{equation}

According to the Wiedemann-Franz law (\ref{eq1}) the electronic part of the thermal conductivity $\kappa_e$ and the electrical conductivity $\sigma$ are closely related to each other, where $L \approx 2.44 \times 10^{-8} \ \mathrm{W \Omega K^{-2}} $ is the Lorenz number. Therefore, the electronic and lattice contributions to the thermal conductivity $\kappa$ can be separated, as shown in table \ref{tab1}. The electronic contribution to the thermal conductivity $\kappa_e$ increases with increasing Sb content. The calculated lattice contribution of the thermal conductivity $\kappa_l$ has a maximum around \textit{x} = 0.75. The lowest lattice contributions, which are best for obtaining the most efficient thermoelectric materials, are therefore found close to the end members (\textit{x} = 0 and 1).

\begin{table}[htbp]
  \centering
 
    \begin{tabular}{l|c|c|c|c|c}
    $T = \mathrm{295 \ K}$ & \ce{GeBi4Te7} & \ce{GeBi3SbTe7} & \ce{GeBi2Sb2Te7} & \ce{GeBiSb3Te7} & \ce{GeSb4Te7} \\
    \toprule
    $\kappa_{tot} \ \mathrm{(W m^{-1} K^{-1})}$     & 0.31 & 0.34 & 0.43 & 0.60 & 0.82 \\
    $\kappa_e \ \mathrm{(W m^{-1} K^{-1})}$ & 0.08 & 0.04 & 0.06 & 0.12 & 0.55 \\
    $\kappa_l \ \mathrm{(W m^{-1} K^{-1})}$ & 0.23 & 0.30 & 0.37 & 0.48 & 0.26 \\
    \bottomrule
    \end{tabular}%
     \caption{Thermal conductivities at 295 K for \textit{x} = 0, 0.25, 0.5 and 1, where $\kappa_l$ is estimated from the measured $\kappa_{tot}$ and $\kappa_e$ using the Wiedemann-Franz law.}
     
  \label{tab1}%
\end{table}%

\begin{figure} [p]
\centering
{\includegraphics[width=\textwidth]{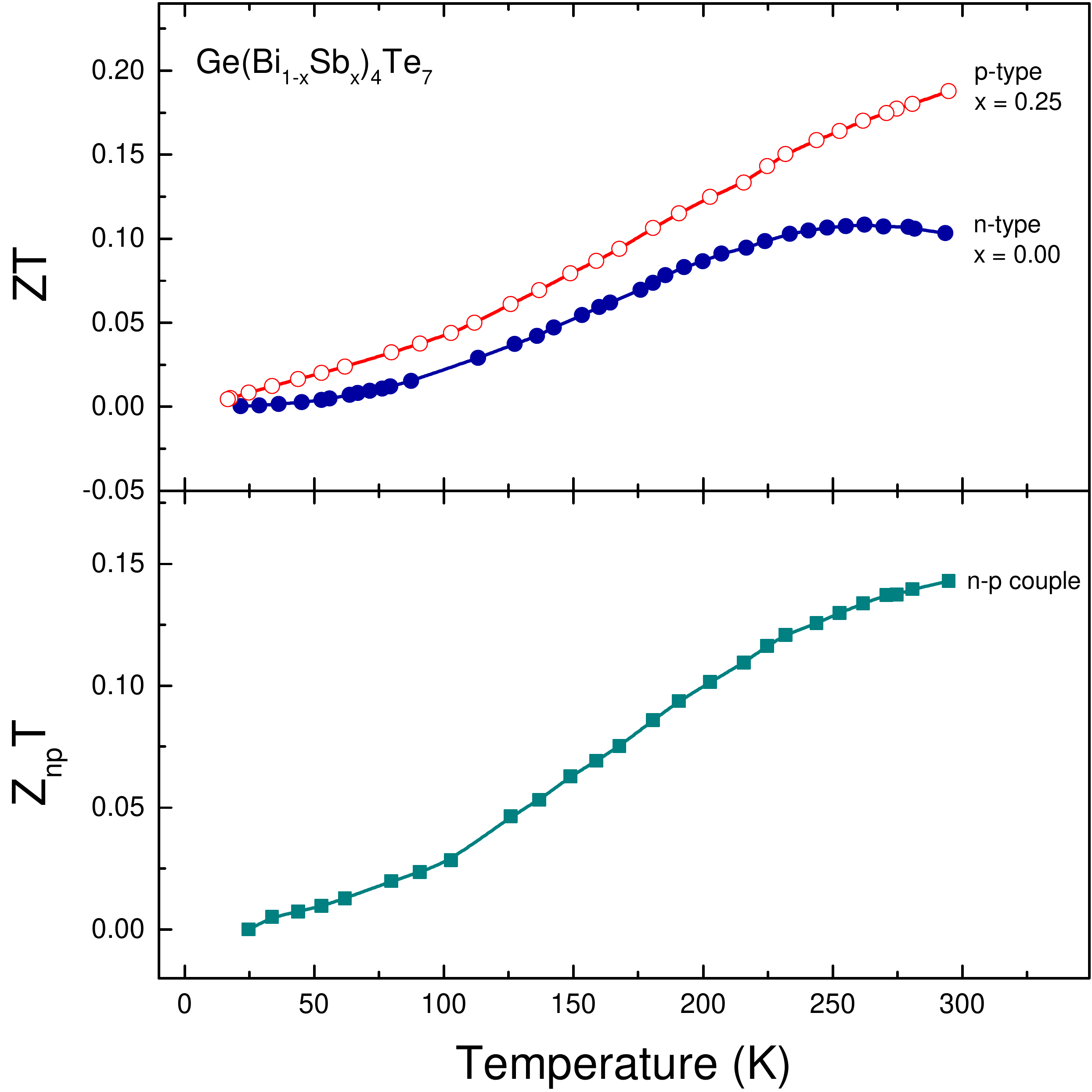}}
\caption{Upper panel: Figure of merit of the most efficient \textit{n}-type (\textit{x} = 0) and \textit{p}-type (\textit{x} = 0.25) samples of the \ce{Ge(Bi_{1-\textit{x}}Sb_{\textit{x}})_4Te7} solid solution in the temperature range from 10 K to 300 K. Lower panel: Thermoelectric efficiency values for the composite \textit{n}-\textit{p} couple, constructed solely from materials in the \ce{Ge(Bi_{1-\textit{x}}Sb_{\textit{x}})_4Te7} solid solution.}
\label{fig:ZT}
\end{figure}

From the combination of thermopower $\alpha$, electrical conductivity $\sigma$ and thermal conductivity $\kappa$ we can calculate the dimensionless thermoelectric figure of merit, $ZT = \alpha^{2} T \sigma \kappa^{-1}$, which is the measure for the efficiency of thermoelectric materials. In the upper panel of figure \ref{fig:ZT}, the figures of merit \textit{ZT} for the most efficient \textit{n}-type and \textit{p}-type materials are illustrated. The most efficient \textit{n}-type material was found for \textit{x} = 0, it reaches a maximal \textit{ZT} of 0.11 around $ T \mathrm{ \ \approx \ 250 \ K}$.  Below \textit{T} = 300 K the \textit{ZT} values for all \textit{n}-type materials are lower than for the most efficient \textit{p}-type compound. The highest \textit{p}-type thermoelectric figure of merit is observed for \textit{x} = 0.25, with a value of \textit{ZT} = 0.19 at room temperature. The data suggests that \textit{ZT} for the \textit{p}-type material may be even larger at higher temperatures than studied here.\\

Our finding of both \textit{n}-type and \textit{p}-type compositions within the solid solution of \ce{Ge(Bi_{1-\textit{x}}Sb_{\textit{x}})_4Te_7} and their reasonably high thermoelectric figure of merit \textit{ZT}, enables the construction of a overall thermoelectric system from this solid solution alone. In the lower panel of figure \ref{fig:ZT} we present the thermoelectric efficiency values for the composite \textit{n}-\textit{p} couple, according to: 

\begin{equation}
Z_{np} T = \frac{\left( \alpha_{p}-\alpha_{n} \right)^{2} \ T}{\left( \left( \frac{\kappa_{p}}{\sigma_{p}} \right)^{\frac{1}{2}} + \left( \frac{\kappa_{n}}{\sigma_{n}} \right)^{\frac{1}{2}} \right)^{2}},
\label{eq2} 
\end{equation}

where $\mathrm{\alpha_{p}, \ \alpha_{n}, \ \sigma_{p}, \ \sigma_{n}, \ and \ \kappa_{p}, \ \kappa_{n}}$ denote thermopowers, electrical conductivities, and thermal conductivities for \textit{p}-type and \textit{n}-type legs, respectively. The composite figure of merit $Z_{np}T$ reaches a value of 0.18 at room temperature for a device with \textit{x} = 0 (\textit{n}-type) and \textit{x} = 0.25 (\textit{p}-type) legs, a value that is clearly lower than the $\mathrm{Z_{np}}T$ values that have been reported for the $\mathrm{(Bi,Sb)_2(Te,Se)_3}$ system \cite{ThuH}, in the temperature regime investigated here.\\

\begin{figure} [p]
\centering
{\includegraphics[width=0.7\textwidth]{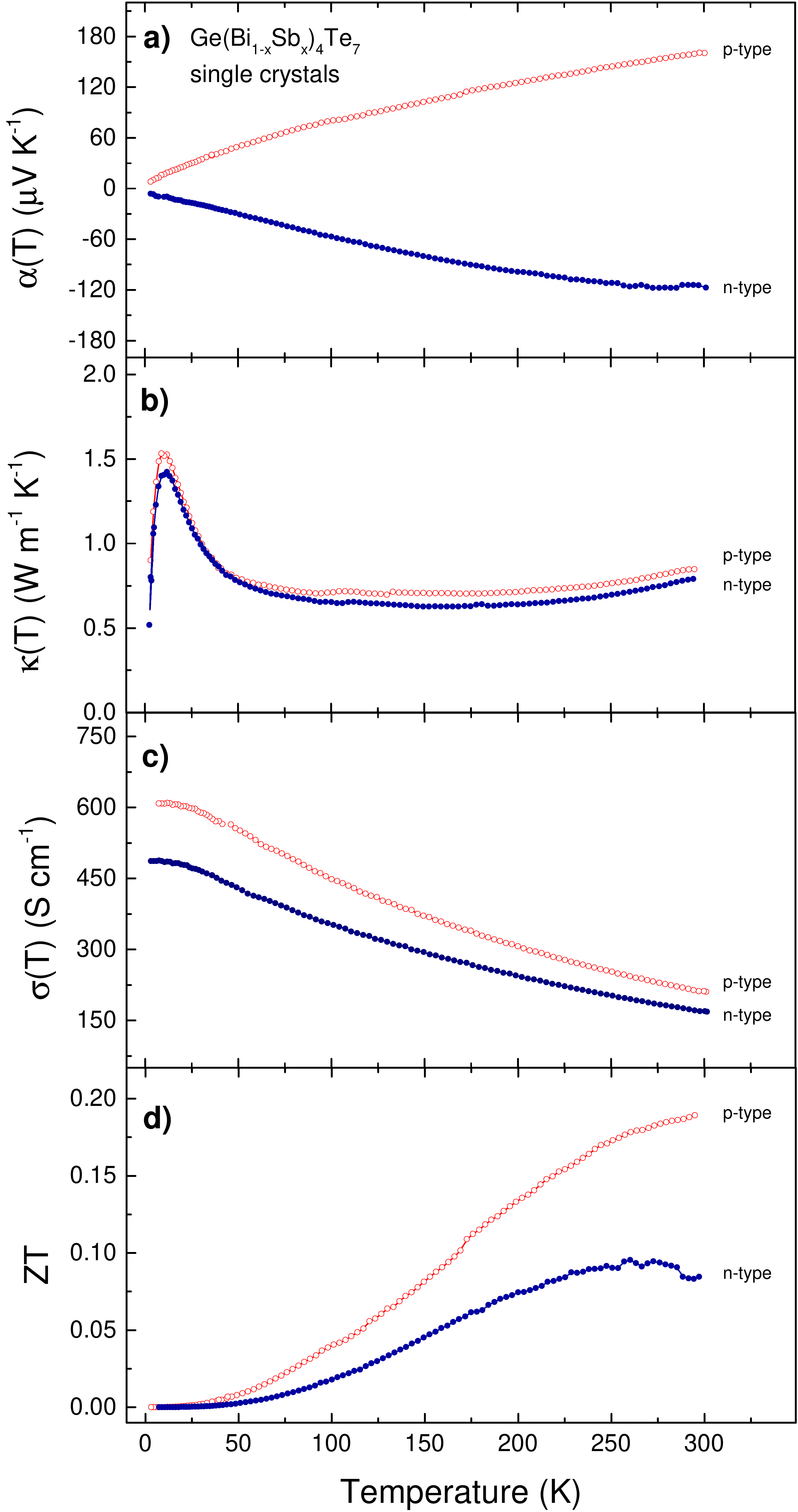}}
\caption{(a)-(c) Temperature dependent measurements between 10 and 300 K of the thermopower $\alpha(T)$, the thermal conductivity $\kappa(T)$ and the electrical conductivity $\sigma(T)$ of single crystals of \textit{x} = 0 (\textit{n}) and 0.25 (\textit{p}). (d) Figure of merit \textit{ZT} of single crystals of \textit{x} = 0 (\textit{n}) and 0.25 (\textit{p}).}
\label{fig:singlecryst}
\end{figure}

The thermopowers $\alpha(T)$, thermal conductivities $\kappa(T)$ and electrical conductivities $\sigma(T)$ were also measured on single crystals of the most efficient thermoelectric materials of the \ce{Ge(Bi_{1-\textit{x}}Sb_{\textit{x}})_4Te_7} solid solution (\textit{x} = 0 and 0.25) between 10 and 300 K, where the respective heat and electrical currents were applied along the \textit{c}-plane. The results of the measurements are presented in figure \ref{fig:singlecryst}. In comparison to the polycrystalline samples, the thermopowers $\alpha$ are larger for both single crystals at room temperature, with $\alpha_p(295 \ \mathrm{K}) = 160 \ \mathrm{\mu V K^{-1}}$ and $\alpha_n(295 \ \mathrm{K}) = -117 \ \mathrm{\mu V K^{-1}}$, respectively. The respective figures of merit \textit{ZT} of both single crystals are almost the same as for their polycrystalline counterparts.  The calculated electronic and lattice contributions to the thermal conductivities $\kappa$ at 295 K are presented in table \ref{tab2}, along with a comparison of the thermoelectric properties at 295 K of the polycrystalline samples and the single crystals of the same composition. The electrical conductivities are higher for single crystals, and therefore also the electronic contribution of the thermal conductivity is increased accordingly. However, as the total thermal conductivities are more than a factor two larger in single crystals, the lattice contributions are still larger for single crystals than for polycrystalline samples (see table 2), which may reflect the higher perfection of the crystal lattices in the single-crystalline samples. This interpretation is supported by the presence of a marked upturn in $\kappa(T)$ at low temperatures (see figure \ref{fig:singlecryst}b), a feature that is typical for clean samples with long phonon mean-free path and which is absent in corresponding data of polycrystalline material (figure \ref{fig:restherm}b). \\

\begin{table}[h]
  \centering

    \begin{tabular}{l|c|c|c|c|c}
    
     & $\kappa$ & $\kappa_e$ & $\kappa_l$ & $\alpha$ & \textit{ZT} \\
        \textit{T} = 295 K & ($\mathrm{W m^{-1} K^{-1}}$) & ($\mathrm{W m^{-1} K^{-1}}$) & ($\mathrm{W m^{-1} K^{-1}}$) & ($\mathrm{\mu V K^{-1}}$) & \\
    \toprule
    \textbf{\ce{GeBi3SbTe7}} (\textit{x} = 0.25) &&&&& \\
    single crystal & 0.84 & 0.16 & 0.68 & 160   & 0.19 \\
    polycrystal & 0.34 & 0.04 & 0.30 & 130   & 0.20 \\
    \toprule
    \textbf{\ce{GeBi4Te7}} (\textit{x} = 0) &&&&& \\ 
    single crystal & 0.78 & 0.12 & 0.66 & -117  & 0.09\\
    polycrystal & 0.31 & 0.08 & 0.23 & -104  & 0.11 \\
    \bottomrule
    \end{tabular}%
  \caption{The thermal conductivities, thermopowers and figures of merit at room temperature of polycrystalline samples and single crystals for \textit{x} = 0 and 0.25.}
  \label{tab2}%
\end{table}%

We note that in order to obtain homogeneous crystals, we have tested different growth conditions. Crystals grown with a cooling rate slower than 50 $ \mathrm{ ^{\circ}C/h}$ have an inhomogeneous composition and likely also an unequal distribution of the defect density. Annealing the crystals at temperatures too close to the melting point leads to crystal growth by vapor transport at various sites in the quartz tube, causing the same homogeneity problems as described above. (Vapor transport may be an alternative route for obtaining crystals of this material.) Variation of the thermoelectric properties among different crystals cleaved from the same 3 - 4 $\mathrm{cm^3}$ boule were found. However, REM/EDXS measurements using a Zeiss-SUPRA-50-VP revealed a homogeneous distribution of the elements. Therefore we expect that the differences in properties must be due to a distribution of defect densities. Similar findings have been presented recently for this class of materials. \cite{Jia} \\

\section{Conclusion}
We have presented data that shows a systematic crossover from \textit{n}-type to \textit{p}-type as a function of increasing Sb content in the \ce{Ge(Bi $,$ Sb)_{4}Te_7} solid solution, and described a method for the growth of chemically homogeneous crystals. As evidenced, these compounds are reasonably good low temperature \textit{n}-type and \textit{p}-type thermoelectric materials for \textit{x} = 0 and \textit{x} = 0.25, respectively. The crossover from \textit{n}-type to \textit{p}-type is not accompanied by an intermediate semiconducting composition region. All samples show metallic behavior, where the electrical conductivity increases for increasing Sb content. Likewise, the thermal conductivity increases for higher Sb contents. \\

In previous work, thermopowers as large as  -148 $\mathrm{\mu V K^{-1}}$ at room temperature were found for polycrystalline samples of non-stoichiometric $\mathrm{Ge_{1 \pm \delta_{1}} Bi_{4 \pm \delta_{2}} Te_{7 \pm \delta_{3}}}$. \cite{nonstoichiometry} Although resistivities and thermal conductivities have not been reported for those materials, a thermoelectric figure of merit of \textit{ZT} $\approx$ 0.4 for \textit{n}-type and \textit{p}-type compounds in this family may be possible, if we assume similar electrical and thermal conductivities to those we have reported here are also found for the non-stoichiometric \textit{n}-type and \textit{p}-type samples of $\mathrm{Ge_{1 \pm \delta_{1}} (Bi,Sb)_{4 \pm \delta_{2}} Te_{7 \pm \delta_{3}}}$. Our findings may therefore suggest a road to obtain better thermoelectric materials in this system upon further optimization. Finally, careful crystal growth of materials at small Sb content increments in the composition vicinity of the \textit{n}- to \textit{p}- crossover near \textit{x} = 0.15 may yield materials with the bulk semiconducting behavior required for characterization of the transport properties of the topological surface states or the fabrication of experimental devices based on those states.\\

\section{Acknowledgments}
The authors would like to thank Hugo Dil, Stefan Muff, and Michael W\"orle for helpful discussion. FvR acknowledges a scholarship from Forschungskredit UZH, grant no. 57161402. The work at Princeton University was supported by grant AFOSR FA9550-10-1-0533.

\section*{References}
\bibliographystyle{naturemag}
\bibliography{biblio}
\appendix

\end{document}